\documentclass[aps,prd,twocolumn,groupedaddress,showpacs,nofootinbib]{revtex4}
\usepackage{graphicx}

\begin{document}

\title{Beyond the perfect fluid hypothesis for dark energy equation of state}

\author{V.F. Cardone$^1$, C. Tortora$^2$, A. Troisi$^1$, S. Capozziello$^2$}
\thanks{Corresponding author\,: V.F. Cardone, {\tt winny@na.infn.it}}
\affiliation{$^1$Dipartimento di Fisica ``E.R. Caianiello'', Universit\`{a} di Salerno and INFN, Sez. di Napoli, Gruppo Coll. di Salerno, via S. Allende, 84081 - Baronissi (Salerno), Italy \\ $^2$Dipartimento di Scienze Fisiche, Universit\`{a} di Napoli ``Federico II'' and INFN, Sez. di Napoli, Compl. Univ. Monte S. Angelo, Edificio N, Via Cinthia, 80126, Napoli, Italy}

\begin{abstract}

Abandoning the perfect fluid hypothesis, we investigate here the possibility that the dark energy equation of state (EoS) $w$ is a nonlinear function of the energy density $\rho$. To this aim, we consider four different EoS describing classical fluids near thermodynamical critical points and discuss the main features of cosmological models made out of dust matter and a dark energy term with the given EoS. Each model is tested against the data on the dimensionless coordinate distance to Type Ia Supernovae and radio galaxies,  the shift and the acoustic peak parameters and the positions of the first three peaks in the anisotropy spectrum of the comic microwave background radation. We propose a possible interpretation of each model in the framework of scalar field quintessence determining the shape of the self interaction potential $V(\phi)$ that gives rise to each one of the considered thermodynamical EoS. As a general result, we demonstrate that replacing the perfect fluid EoS with more generar expressions gives both the possibility of successfully solving the problem of cosmic acceleration escaping the resort to phantom models.

\end{abstract}


\maketitle

\section{Introduction}

The end of the 21st century has left as unexpected legacy a new picture of the universe depicted as a spatially flat manifold with a subcritical matter content presently undergoing a phase of accelerated expansion. An impressive amount of astrophysical evidences on different scales, from the anisotropy spectrum of cosmic microwave background radiation (hereafter CMBR) \cite{Boom,CMBR,WMAP} to the Type Ia Supernovae (hereafter SNeIa) Hubble diagram \cite{SNeIa,Riess04}, the large scale structure \cite{LSS} and the matter power spectrum determined by the Ly$\alpha$ forest data \cite{Lyalpha}, represent observational cornerstones that put on firm grounds the picture of the universe described above.

Although the classical cosmological constant $\Lambda$ \cite{Lambda} represents the best fit to the full set of observational data \cite{Teg03,Sel04}, the well known {\it coincidence} and {\it fine tuning} problems have lead cosmologists to look for alternative candidates that are collectively referred to as {\it dark energy}. In the most investigated scenario, dark energy originates from a scalar field $\phi$, dubbed {\it quintessence},  running down its self interaction potential $V(\phi)$ so that an effective fluid with negative pressure contributes to the energy budget of the universe (for comprehensive reviews see, for instance, \cite{QuintRev}). It is, however, also possible that dark energy and dark matter are actually two different manifestation of the same substance. In such models, collectively referred to as {\it unified dark energy} (UDE), a single fluid with an exotic equation of state behaves as dark energy at the lowest energy scales and as and dark matter at higher energies \cite{Chaplygin,tachyon,Hobbit}. It is worth remembering that a variant of UDE models has recently been investigated by introducing models which are able to give rise to both inflation and cosmic acceleration \cite{ie,Od} also solving the problem of phantom quintessence \cite{Odphan}. 

Notwithstanding the strong efforts made to solve this puzzle, none of the proposed explanations is fully satisfactory and free of problems. This disturbing situation has motivated much interest toward a radically different approach to the problem of cosmic acceleration. It has therefore been suggested that cosmic speed up is an evidence for the need of {\it new physics} rather than a new fluid. Much interest has then been devoted to models according to which standard matter is the only physical ingredient, while the Friedmann equations are modified, possibly as a consequence of braneworld scenarios \cite{DGP}. In this same framework, higher order theories of gravity represent a valid alternative to the dark energy approach. Also referred to as {\it curvature quintessence}, in these models, the gravity Lagrangian is generalized by replacing the Ricci scalar curvature $R$ with a generic function $f(R)$ so that an effective dark energy\,-\,like fluid appears in the Friedmann equations and drives the accelerated expansion. Different models of this kind have been explored and tested against observations considering the two possible formulations that are obtained adopting the metric \cite{capozcurv,review,noicurv,MetricRn,CCT} or the Palatini \cite{PalRn,lnR,Allemandi} formulation. 

From this overview of the different theoretical models proposed so far, it is clear that rather little is definitively known on the nature and the fundamental properties of the dark energy even if some model independent constraints on its present day value and on its first derivative may be inferred from nonparametric analyses (see, e.g., \cite{nonparam}). It is worth noting, however, that, in all the models considered so far (with the remarkable exception of UDE models), it has been aprioristically assumed that dark energy behaves as a perfect fluid so that its EoS is linear in the energy density. Actually, from elementary thermodynamics, we know that a real fluid is never perfect \cite{Rowlinson} and, on the contrary, such an assumption is more and more inadequate as the fluid approaches its thermodynamical {\it critical points} or during phase transitions. Given our fundamental ignorance about the properties of the dark sector, we cannot exclude the possibility that the universe is in a sort of critical point so that its constituents cannot be treated as perfect fluids. While the matter term, whatever its nature, may be safely modelled as a dustlike component (i.e., its EoS simplifies to $p = 0$), forcing the dark energy to be a perfect fluid is a rough simplification that may lead to neglect the impact on the dynamics of its true properties. Moreover, such an unmotivated approach could lead to systematically wrong results and hence to misleading inferences on the dark energy nature. Motivated by these considerations, it is therefore worth exploring what are the consequences of abandoning the perfect fluid EoS. A first step in this direction has been performed by Capozziello et al. \cite{vdw} who have considered a model in which a single fluid with a Van der Waals EoS accounts for both dark matter and dark energy (see also \cite{Kr03}). From classical thermodynamics, we know that the Van der Waals EoS is best suited to describe the behaviour of real gas with a particular attention to the phase transitions phenomena. Actually, the Van der Waals EoS is only one of the possible choices in these regimes. Elaborating further on the idea put forward by Capozziello et al., we explore here other thermodynamical EoS all sharing the properties of having been proposed to work well also for fluids in critical conditions. Moreover, these EoS contain the perfect fluid EoS as a limiting case thus representing useful and more realistic generalizations. It is worth stressing that such an approach better reflects our ignorance of the dark energy nature and should prevent us from deducing theoretically biased conclusions on its nature. 

The plan of the paper is as follows. In Sect.\,II, we present the EoS considered giving their expressions and characterizing parameters. The dynamics of cosmological models comprising dust matter and a dark energy fluid with such an EoS is discussed in Sect.\,III where we determine the redshift evolution of the main physical quantities of interest. Matching with the data allows to investigate the viability of the different EoS and constrain their parameters. The method we use and the results we get are presented in Sect.\,IV, while the position of the peaks in the CMBR anisotropy spectrum is evaluated in Sect.\,V and compared with the WMAP determination. In Sect.\,VI, we reinterpret the models proposed in the framework of scalar field quintessence determining the self interaction potential that gives rise to a dark energy model with the given EoS. Conclusions are presented in Sect.\,VII, while in Appendix A we give some further details on the EoS from the thermodynamic point of view. 

\section{Equations of state}

The dynamical system describing a Friedmann--Robertson--Walker (FRW) cosmology is given by the Friedmann equations \cite{CosmoBooks}\,:

\begin{equation}
\frac{\ddot{a}}{a} = - \frac{4 \pi G}{3} \ (\rho_M + \rho_X + 3 p_X) \ , 
\label{eq: fried1}
\end{equation}

\begin{equation}
H^2 = \frac{8 \pi G}{3} (\rho_M + \rho_X) \ , 
\label{eq: fried2}
\end{equation}
and the continuity equations for each of the two fluids\,:

\begin{equation}
\dot{\rho_i} + 3 H \ (\rho_i + p_i) = 0 \ , 
\label{eq: continuity}
\end{equation}
where $a$ is the scale factor, $H = \dot{a}/a$ the Hubble parameter, the dot denotes the derivative with respect to cosmic time and we have assumed a spatially flat universe in agreement with what is inferred from CMBR anisotropy spectrum \cite{Boom,CMBR,WMAP}. Eqs.(\ref{eq: fried1}), (\ref{eq: fried2}) and (\ref{eq: continuity}) are derived by the Einstein field equations and the contracted Bianchi identities assuming that the source of the gravitational field is a a mixture of matter with energy density $\rho_M$ and pressure $p_M = 0$ and an additional negative pressure fluid (which is usually referred to as {\it dark energy}) with energy density $\rho_X$ and pressure $p_X$. To close the system and determine the evolution of the scale factor $a$ and of the other quantities of interest, the equation of state (hereafter EoS) of the dark energy fluid (i.e. a relation between $\rho_X$ and $p_X$) is needed.  

Unfortunately, this is a daunting task given our complete ignorance of the dark energy nature and of its fundamental properties. Motivated by the discussion in the introduction, we explore here some EoS all sharing the properties of working well even when the fluid is near critical points or phase transitions. A textbook example is the Van der Waals EoS\,:
\begin{equation}
p_X = \frac{\gamma \rho_X}{1 - \beta \rho_X} - \alpha \rho_X^2 \ ,
\label{eq: vdweq}
\end{equation}
wherec $\alpha$ and $\beta$, in the thermodynamics analogy, may be related to limiting values of the pressure and the volume, while $\gamma$ is the usual barotropic factor. The Van der Waals EoS reduces to the perfect fluid case in the limit $\alpha,\beta \rightarrow 0$. The dynamics of the corresponding cosmological model has been yet investigated both in the framework of UDE models \cite{vdw} and as a dark energy fluid \cite{Kr03} so that we do not consider it again here. On the other hand, there are other EoS that are worth investigating. However, we limit our attention to EoS described by two parameters only in order to both narrow the parameter space to explore and avoid introducing too large a degeneracy among the quantities we have to determine.  

Let us first define $\eta(z)$ and $\tilde{p}(z)$ as $\rho_X(z)/\rho_{crit}$ and  $p_X(z)/\rho_{crit}$ respectively, being $\rho_{crit} = 3 H_0^2/8 \pi G$ the present day critical density of the universe. The EoS may then be evaluated as $w = p_X/\rho_X = \tilde{p}/\eta$. For the different models we consider, $w$ is a non linear function of $\eta$ and is given as follows.

\begin{enumerate}

\item[i.]{{\it Redlich\,-\,Kwong}\,:

\begin{equation}
w_{RK} = \beta \times \frac{1 - \sqrt{3 - 2 \sqrt{2}} \alpha \eta}{1 - (1 - \sqrt{2}) \alpha \eta} \ .
\label{eq: wrk}
\end{equation}}

\item[ii.]{{\it Modified Berthelot}\,:

\begin{equation}
w_{MB} = \frac{\beta}{1 + \alpha \eta} \ .
\label{eq: wmb}
\end{equation}}

\item[iii.]{{\it Dieterici}\,:

\begin{equation}
w_{Dt} = \frac{\beta \exp{[ 2 (1 - \alpha \eta)]}}{2 - \alpha \eta} \ .
\label{eq: wdt}
\end{equation}}

\item[iv.]{{\it Peng\,-\,Robinson}\,:

\begin{equation}
w_{PR} = \frac{\beta}{1 - \alpha \eta} \times \left [ 1 - 
\frac{(c_a/c_b) \alpha \eta}{(1 + \alpha \eta)/(1 - \alpha \eta) + \alpha \eta} \right ]
\label{eq: wpr}
\end{equation}
with $c_a \simeq 1.487$ and $c_b \simeq 0.253$.}

\end{enumerate}
In Eqs.(\ref{eq: wrk})\,-\,(\ref{eq: wpr}), the two parameters $\alpha$ and $\beta$ are related to the critical values of density and pressure of the fluid. In particular, for all cases, $\alpha \propto \rho_{crit}/\rho_c$, while $\beta \propto p_c/\rho_c$ with $\rho_c$ and $p_c$ the values of the energy density and pressure respectively at the critical point of the fluid\footnote{See the Appendix for the definition of critical points and the exact expression of $\alpha$ and $\beta$.}. Note that, for $\alpha = 0$, all the EoS above reduces to $w = cst$, i.e. to the perfect fluid one. The condition $\alpha = 0$ is achieved for an infinite critical density which means that the fluid has no critical points. This is indeed the case of the perfect fluid and is the reason why such a description is highly unrealistic. 

It is convenient, however, in the application to express these two parameters in terms of more handable and meaningful quantities. To this aim, we first define

\begin{equation}
y \equiv \alpha \eta(z = 0) = \alpha (1 - \Omega_M) 
\Rightarrow \alpha = y/(1- \Omega_M)
\label{eq: defypar} 
\end{equation}
where we have used the flatness condition $\Omega_M + \Omega_X = 1$. Combining Eqs.(\ref{eq: fried1}) and (\ref{eq: fried2}), using the definition of deceleration parameter $q \equiv - a \ddot{a}/\dot{a}^2$ and evaluating the result at $z = 0$, we get the well known relation\,:

\begin{equation}
q_0 = \frac{1}{2} + \frac{3}{2} \Omega_X w_0 \ .
\label{eq: qz}
\end{equation}
Introducing one of Eqs.(\ref{eq: wrk})\,-\,(\ref{eq: wpr}) into Eq.(\ref{eq: qz}) with the condition $\eta(z = 0) = \Omega_X$, we can express $\beta$ in terms of $q_0$, $y$ and $\Omega_M$ thus obtaining what follows.

\begin{enumerate}

\item[i.]{{\it Redlich\,-\,Kwong}\,:

\begin{equation}
\beta = \frac{(2 q_0 - 1)[1 - (1 - \sqrt{2}) y]}{3 (1 - \Omega_M)(1 - \sqrt{3 - 2 \sqrt{2}} y)} \ .
\label{eq: brk}
\end{equation}}

\item[ii.]{{\it Modified Berthelot}\,:

\begin{equation}
\beta = \frac{(2 q_0 - 1)(1 + y)}{3 (1 - \Omega_M)}  
\label{eq: bmb}
\end{equation}}

\item[iii.]{{\it Dieterici}\,:

\begin{equation}
\beta = \frac{(2 q_0 - 1)(2 - y) \exp{[ -2 (1 - y)]}}{3 (1 - \Omega_M)} \ .
\label{eq: bdt}
\end{equation}}

\item[iv.]{{\it Peng\,-\,Robinson}\,:

\begin{equation}
\beta = - \frac{c_b (2 q_0 - 1)(y - 1)(y^2 - 2y - 1)}
{3 (1 - \Omega_M) [ c_a y (1 - y) + c_b (y^2 - 2y - 1)]} \ .
\label{eq: bpr}
\end{equation}}

\end{enumerate}
Note that it is $y$ rather than $\alpha$ to determine $\beta$ so that it is this parameter that will be constrained by the fitting procedure. Moreover, $q_0$ and $\Omega_M$ are more familiar quantities than $\beta$ so that it is easier to choose intervals over which they can take values.

\section{Redshift evolution}

\begin{figure}
\centering \resizebox{8.5cm}{!}{\includegraphics{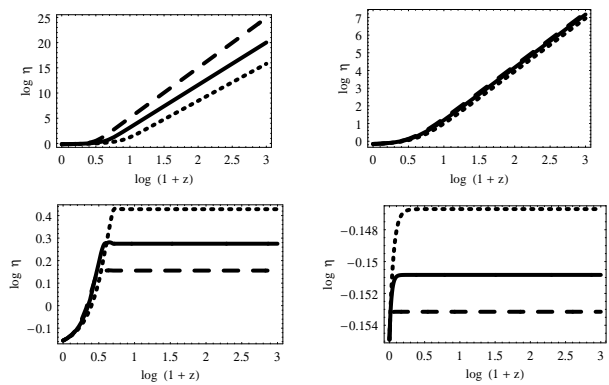}}
\caption{The evolution against the redshift of the dimensionless energy density $\eta(z)$ for the $RK$ (upper left), $MB$ (upper right), $Dt$ (lower left) and $PR$ (lower right) EoS respectively. For all models, we set $(q_0, \Omega_M) = (-0.5, 0.3)$. Short dashed, solid and long dashed lines refer to different values of $y$, namely $y = 0.5, 0.75, 1.0$ for the $RK$ model, $y = 1.5, 2.5, 3.5$ for the $MB$ one, $y = 0.5, 0.7, 0.9$ for the $Dt$ and $PR$ models.}
\label{fig: ez}
\end{figure}

For a given EoS, it is possible to determine how the main physical quantities (such as the energy density, the EoS and the Hubble parameter) evolves with the redshift $z$. To this aim, we first change variable from $t$ to $z$ in the continuity equation (\ref{eq: continuity}) which thus is rewritten as\,:

\begin{equation}
\frac{d\eta}{dz} = \frac{3 (1 + w) \eta(z)}{1 + z}
\label{eq: cont}
\end{equation}
with the initial condition $\eta(z = 0) = \Omega_X$. Note that, for the EoS we are considering, Eq.(\ref{eq: cont}) is a first order nonlinear differential equation that cannot be solved analytically. However, a numerical integration is straightforward provided that the parameters $(q_0, y, \Omega_M)$ are given. Fig.\,\ref{fig: ez} shows the results for the different EoS described in the previous section for some illustrative set of parameters. Note that, hereon, to shorten the notation, we will use the acronyms $RK$, $MB$, $Dt$ and $PR$ referring to the Redlich\,-\,Kwong, Modified Berthelot, Dieterici and Peng\,-\,Robinson cases respectively.

As it is clearly shown, not surprisingly, different EoS may lead to radically different evolutions for the energy density. This is particularly true comparing the upper panels with the lower ones. For the $RK$ and $MB$ cases, $\log{\eta}$ is an almost linear increasing function of $z$, i.e. $\eta(z)$ has an approximately power\,-\,law like decrease with the cosmic time $t$ over a large range. As a consequence, in the far past, the dark energy component does not fade away, but still contributes to the energy budget during the usually matter dominated epoch. This behaviour may be problematic for the impact on structure formation and nucleosynthesis. For the $RK$ case, this problem may be particularly worrisome since, as can be inferred from Fig.\,\ref{fig: ez}, for high $z$, $\eta \sim (1 + z)^{\gamma}$ with $\gamma$ larger than $3$ for some combination of the parameters $(q_0, y, \Omega_M)$. The situation is less dramatic for the $MB$ EoS since, although we still get $\eta \sim (1 + z)^{\gamma}$ for high $z$, now $\gamma$ is smaller than $3$ so that the dark energy term becomes quite small during the matter dominated era. Note also that the evolution of $\eta(z)$ only weakly depends on $y$ for the $MB$ model thus suggesting a serious degeneracy in this quantity. The situation is radically different for the $Dt$ and $PR$ models. As Fig.\,\ref{fig: ez} shows, in these cases, $\eta(z)$ quickly approaches a constant value so that, in the past, the dark energy component does not disappear, but plays the same role as a cosmological constant term. Note, in particular, that this regime is achieved very soon for the $PR$ EoS in which case $\eta(z)$ is almost constant for quite small values of $z$. These results are reassuring since the energy density of the dark energy component becomes vanishingly small with respect to that of the matter during both the structure formation and nucleosynthesis epochs so that we are quite confident that these processes are not altered by the use of unusual EoS.     

\begin{figure}
\centering \resizebox{8.5cm}{!}{\includegraphics{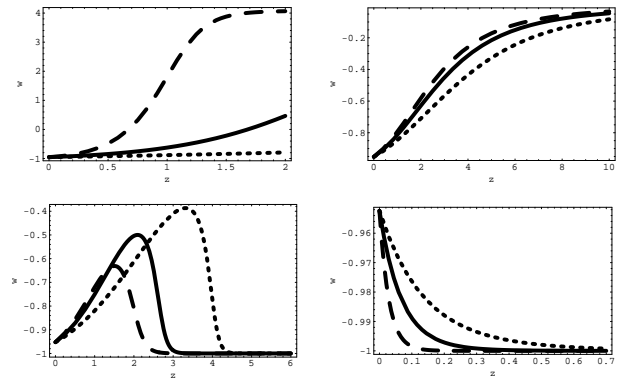}}
\caption{The evolution against the redshift of the EoS parameter $w(z)$ for the $RK$ (upper left), $MB$ (upper right), $Dt$ (lower left) and $PR$ (lower right) EoS respectively. Model parameters are set as in Fig.\,\ref{fig: ez}.}
\label{fig: wz}
\end{figure}

Having determined $\eta(z)$, it is now straightforward to compute how the EoS depend on the redshift. The result is shown in Fig.\,\ref{fig: wz} where we plot $w(z)$ for the different models adopting for the model parameters the same values as in Fig.\,\ref{fig: ez}. Although the behaviour of $\eta(z)$ is qualitatively similar in some cases, the shape of $w(z)$ is radically different for the models we are considering so that we discuss them separately.

First, let us consider the $RK$ EoS. For the models in the upper left panel of Fig.\,\ref{fig: wz}, the EoS turns out to be an increasing function of the redshift $z$ with the largest value of $y$ leading to higher $w$ at high $z$. In particular, $w$ may become positive for sufficiently large values of $y$. However, we have checked that this result strongly depends on $\Omega_M$. Actually, for values of $\Omega_M \ge 0.35$, the EoS becomes more and more negative as $z$ increases so that the fluid behaves as kind of {\it superphantom}. As a general rule, however, we stress that, for $y \le 1$, $w(z \simeq 0) \simeq -1$ so that in the present epoch the EoS mimics that of the cosmological constant. 

The $MB$ EoS is shown in the upper right panel of Fig.\,\ref{fig: wz}, but the main trend should be inferred directly from Eq.(\ref{eq: wmb}). Since $\eta(z)$ is an increasing function of $z$, it is easy to understand that, whatever are the values of $(q_0, y, \Omega_M)$, $w(z)$ will vanish in the past so that a dust\,-\,like EoS is asymptotically achieved. Note, however, that the convergence may be quite slow depending on $y$\,: the larger is $y$, the higher is $w$ at a given $z$ so that the quicker is the convergence toward the asymptotic null value. Given this behaviour, it is tempting to use the $MB$ EoS as a proposal for a UDE model. From the point of view of the parameters, this model may be obtained imposing $\Omega_X = 1 - \Omega_b$ with $\Omega_b$ the baryon density parameter. However, we prefer to not fix $\Omega_X$ and determine it later from matching with the data.

Let us consider now $w(z)$ for the $Dt$ parametrization (lower left panel in Fig.\,\ref{fig: wz}). In contrast with the other cases considered, $w(z)$ is not a monotonic function of the redshift, but it has rather an asymmetric bell\,-\,shaped behaviour. In particular, the height of the peak is larger for smaller values of $y$ and its position shifts toward right (i.e., larger values of $z$) with the increasing of $y$. The most remarkable feature is, however, the asymptotic approach toward the cosmological constant value $w = -1$ that is achieved later for smaller $y$. A similar behaviour is consistent with the result shown in Fig.\,\ref{fig: ez} according to which $\eta(z)$ becomes constant for values of $z$ sufficiently high. Comparing the two plots, we see that $\eta(z)$ starts being approximately constant as soon as $w(z)$ is indistinguishable from $-1$ so that everything works as for the cosmological constant. Even if not shown in the plot, we note that $w(z)$ approaches $-1$ also in the limit $z \rightarrow -1$, i.e. in the asymptotic future, so that a de Sitter like expansion is achieved.

Finally, let us discuss the case of the $PR$ EoS which is depicted in the lower right panel of Fig.\,\ref{fig: wz}. As a striking result, we get that $w(z)$ starts from a value very close to -1 and very soon reaches $w = -1$ after which it does not change anymore. This behaviour nicely explains why $\eta(z)$ is approximately constant over almost the full evolutionary history. It is also worth noting that, although $w(z)$ depends significantly on $y$, the numerical change in its value is too small to be detected. As a consequence, it is likely that matching with the data will be unable to efficiently constrain this parameter. 

Another interesting dynamical quantity is the deceleration parameter $q$. Combining the Friedmann equations, it is straightforward to get\,:

\begin{equation}
q(z) = \frac{1}{2} + \frac{3}{2} \frac{w(z) \eta(z)}{\Omega_M (1 + z)^3 + \eta(z)}
\label{eq: qvsz}
\end{equation}
so that, having yet evaluated both $\eta(z)$ and $w(z)$, it is immediate to compute $q(z)$. It turns out that, for all the EoS considered, the evolution of $q(z)$ is quite similar over the redshift range probed by the most of the available data. Moreover, it is remarkable that there is almost no dependence at all on $y$ for the $MB$, $Dt$ and $PR$ models, while a weak dependence is present in the case of the $RK$ EoS. In this latter case, it is important to stress that the adopted value of $\Omega_M$ plays a fundamental role with values of $\Omega_M \ge 0.35$ leading to $q(z) < 0$ for all $z > 0$, i.e. these models are never decelerating. For all other cases, instead, $q(z)$ changes sign so that the transition redshift, defined as $q(z_T) = 0$, turns out to be positive in agreement with some recent estimates.

As a final issue, we have also numerically evaluated how the scale factor depends on the cosmic time $t$. Introducing the normalized time variable $\tau = t/t_0$ (with $t_0$ the present age of the universe), it turns out that $a(\tau)$ is almost linear over the most of the universe evolution and, what is more interesting, is independent of $y$. Actually, this is only a result of having used the dimensionless time $\tau$ rather than the physical time $t$. Since, as we have checked, $t_0$ depends significantly on the combination of the model parameters $(q_0, y, \Omega_M)$, transforming from $a(\tau)$ to $a(t)$ introduce the expected dependence on $y$. As a general result, $a(t)$ does not diverge in any finite time so that any {\it Big Rip} is avoided even if $w$ may lie today in the phantom ($w < -1$) regime.

\section{Constraining the EoS}

The discussion above has shown that the dynamics of a cosmological model filled with dust matter and a dark energy fluid whose EoS is one of those proposed in Sect.\,II is reasonable and not affected by any pathological behaviour (provided the parameters are chosen with some care in the case of the $RK$ model). It is therefore interesting to compare the models with the available observations in order to both investigate the viability of the model itself and constrain its parameters.

\subsection{The method}

In order to constrain the EoS characterizing parameters, we maximize the following likelihood function\,:

\begin{equation}
{\cal{L}} \propto \exp{\left [ - \frac{\chi^2({\bf p})}{2} \right ]}
\label{eq: deflike}
\end{equation}
where {\bf p} denotes the set of model parameters and the pseudo\,-\,$\chi^2$ merit function is defined as\,:

\begin{eqnarray}
\chi^2({\bf p}) & = & \sum_{i = 1}^{N}{\left [ \frac{r^{th}(z_i, {\bf p}) - r_i^{obs}}{\sigma_i} \right ]^2} \nonumber \\
~ & + & 
\displaystyle{\left [ \frac{{\cal{R}}({\bf p}) - 1.716}{0.062} \right ]^2} 
+ \displaystyle{\left [ \frac{{\cal{A}}({\bf p}) - 0.469}{0.017} \right ]^2}  \ .
\label{eq: defchi}\
\end{eqnarray}
Let us discuss briefly the different terms entering Eq.(\ref{eq: defchi}). In the first one, we consider the dimensionless coordinate distance $y$ to an object at redshift $z$ defined as\,:

\begin{equation}
r(z) = \int_{0}^{z}{\frac{dz'}{E(z')}} 
\label{eq: defy}
\end{equation}
and related to the usual luminosity distance $D_L$ as $D_L = (1 + z) r(z)$. Daly \& Djorgovki \cite{DD04} have compiled a sample comprising data on $y(z)$ for the 157 SNeIa in the Riess et al. \cite{Riess04} Gold dataset and 20 radiogalaxies from \cite{RGdata}, summarized in Tables\,1 and 2 of \cite{DD04}. As a preliminary step, they have fitted the linear Hubble law to a large set of low redshift ($z < 0.1$) SNeIa thus obtaining\,:

\begin{displaymath}
h = 0.664 \pm 0.008 \ . 
\end{displaymath}
We thus set $h = 0.664$ in order to be consistent with their work, but we have checked that varying $h$ in the $68\%$ CL quoted above does not alter the main results. Furthermore, the value we are using is consistent also with $H_0 = 72 \pm 8 \ {\rm km \ s^{-1} \ Mpc^{-1}}$ given by the HST Key project \cite{Freedman} based on the local distance ladder and with the estimates coming from the time delay in multiply imaged quasars \cite{H0lens} and the Sunyaev\,-\,Zel'dovich effect in X\,-\,ray emitting clusters \cite{H0SZ}.

The second term in Eq.(\ref{eq: defchi}) makes it possible to extend the redshift range over which $y(z)$ is probed resorting to  the distance to the last scattering surface. Actually, what can be determined from the CMBR anisotropy spectrum is the so called {\it shift parameter} defined as \cite{WM04,WT04}\,:

\begin{equation}
R \equiv \sqrt{\Omega_M} y(z_{ls})
\label{eq: defshift}
\end{equation}
where $z_{ls}$ is the redshift of the last scattering surface which can be approximated as \cite{HS96},:

\begin{equation}
z_{ls} = 1048 \left ( 1 + 0.00124 \omega_b^{-0.738} \right ) 
\left ( 1 + g_1 \omega_M^{g_2} \right )
\label{eq: zls}
\end{equation}
with $\omega_i = \Omega_i h^2$ (with $i = b, M$ for baryons and total matter respectively) and $(g_1, g_2)$ given in Ref.\,\cite{HS96}. The parameter $\omega_b$ is well constrained by the baryogenesis calculations contrasted to the observed abundances of primordial elements. Using this method, Kirkman et al. \cite{Kirk} have determined\,:

\begin{displaymath}
\omega_b = 0.0214 \pm 0.0020 \ .
\end{displaymath}
Neglecting the small error, we thus set $\omega_b = 0.0214$ and use this value to determine $z_{ls}$. It is worth noting, however, that the exact value of $z_{ls}$ has a negligible impact on the results and setting $z_{ls} = 1100$ does not change none of the constraints on the other model parameters.

Finally, the third term in the definition of $\chi^2$ takes into account the recent measurements of the {\it acoustic peak} in the large scale correlation function at $100 \ h^{-1} \ {\rm Mpc}$ separation detected by Eisenstein et al. \cite{Eis05} using a sample of 46748 luminous red galaxies (LRG) selected from the SDSS Main Sample \cite{SDSSMain}. Actually, rather than the position of acoustic peak itself, a closely related quantity is better constrained from these data defined as \cite{Eis05}\,:

\begin{equation}
{\cal{A}} = \frac{\sqrt{\Omega_M}}{z_{LRG}} \left [ \frac{z_{LRG}}{E(z_{LRG})}
y^2(z_{LRG}) \right ]^{1/3}
\label{eq: defapar}
\end{equation}
with $z_{LRG} = 0.35$ the effective redshift of the LRG sample. As it is clear, the ${\cal{A}}$ parameter depends not only on the dimensionless coordinate distance (and thus on the integrated expansion rate), but also on $\Omega_M$ and $E(z)$ explicitly which removes some of the degeneracies intrinsic in distance fitting methods. Therefore, it is particularly interesting to include ${\cal{A}}$ as a further constraint on the model parameters using its measured value \cite{Eis05}\,:

\begin{displaymath}
{\cal{A}} = 0.469 \pm 0.017 \ .
\end{displaymath}
Note that, although similar to the usual reduced $\chi^2$ introduced in statistics, the reduced $\chi^2$ (i.e., the ratio between the $\chi^2$ and the number of degrees of freedom) is not forced to be 1 for the best fit model because of the presence of the priors on ${\cal{R}}$ and ${\cal{A}}$ and since the uncertainties $\sigma_i$ are not Gaussian distributed, but take care of both statistical errors and systematic uncertainties. With the definition (\ref{eq: deflike}) of the likelihood function, the best fit model parameters are those that maximize ${\cal{L}}({\bf p})$. However, to constrain a given parameter $p_i$, one resorts to the marginalized likelihood function defined as\,:

\begin{equation}
{\cal{L}}_{p_i}(p_i) \propto \int{dp_1 \ldots \int{dp_{i - 1} \int{dp_{i + 1} ... \int{dp_n {\cal{L}}({\bf p})}}}}
\label{eq: defmarglike}
\end{equation}
that is normalized at unity at maximum. Denoting with $\chi_0^2$ is the value of the $\chi^2$ for the best fit model, the $1 \ {\rm and} \ 2 \sigma$ confidence regions are determined by imposing $\Delta \chi^2 = \chi^2 - \chi_0^2 = 1$ and $\Delta \chi^2 = 4$ respectively.

\subsection{Results}

We have applied the likelihood analysis described above to the four EoS presented in Sect.\,II obtaining constraints on the model parameters $(q_0, y, \Omega_M)$. The results obtained are summarized in Table\,I where we give the best fit values and $68\%$ and $95\%$ confidence ranges for each parameter. Note that the range tested for $y$ is set on a case by case basis as discussed in the following. Given $(q_0, y, \Omega_M)$ for an EoS, we may also evaluate some other interesting physical quantities. Since the uncertainties on the model parameters are not Gaussian distributed, a naive propagation of the errors is not possible. Moreover, we have not an analytical expression for some quantities such as, e.g., the transition redshift. We thus estimate the 68$\%$ and 95$\%$ confidence ranges on the derived quantities by randomly generating 20000 points $(q_0, y, \Omega_M)$ using the marginalized likelihood functions of each parameter and then deriving the likelihood function of the corresponding quantity. Although not statistically well motivated, this procedure gives a conservative estimate of the uncertainties which is enough for our aims. 

\begin{table*}
\begin{center}
\begin{tabular}{|c|c|c|c|c|c|c|c|}
\hline
Id & $q_0$ & $y$ & $\Omega_M$ & ${\cal{A}}$ & ${\cal{R}}$ & $z_T$ & $t_0$ (Gyr)  \\
\hline  \hline
~ & ~ & ~ & ~ & ~ & ~ & ~ & ~ \\

$RK$ & $-0.490_{-0.005 -0.009}^{+0.005 +0.009}$ & $1.90_{-0.02 -0.04}^{+0.02 +0.04}$ & $0.355_{-0.001 -0.002}^{+0.001 +0.003}$ & --- & --- & --- & --- \\
~ & ~ & ~ & ~ & ~ & ~ & ~ & ~ \\

$MB$ & $-0.55_{-0.04 -0.09}^{+0.04 +0.09}$ & $0 \ (\le 4.2 \ \le 6.0)$ & $0.28_{-0.02 -0.04}^{+0.02 +0.04}$ & $0.474_{-0.018 -0.033}^{+0.014 +0.030}$ & $1.738_{-0.032 -0.095}^{+0.022 +0.042}$ & $0.692_{-0.042 -0.082}^{+0.050 +0.094}$ & $14.29_{-0.21 -0.41}^{+0.22 +0.43}$ \\

~ & ~ & ~ & ~ & ~ & ~ & ~ & ~ \\

$Dt$ & $-0.55_{-0.04 -0.09}^{+0.04 +0.09}$ & $0.9 \ (\le 0.9)$ & $0.28_{-0.02 -0.04}^{+0.02 +0.04}$ & $0.471_{-0.016 -0.033}^{+0.015 +0.033}$ & $1.733_{-0.034 -0.065}^{+0.021 +0.045}$ & $0.706_{-0.062 -0.132}^{+0.053 +0.107}$ & $14.32_{-0.24 -0.49}^{+0.23 +0.49}$ \\
~ & ~ & ~ & ~ & ~ & ~ & ~ & ~ \\

$PR$ & $-0.55_{-0.07 -0.46}^{+0.07 +0.32}$ & $0.9 \ (\ge 0.57 \ \ge 0.24)$ & $0.28_{-0.01 -0.03}^{+0.02 +0.04}$ & $0.476_{-0.012 -0.025}^{+0.015 +0.030}$ & $1.742_{-0.012 -0.027}^{+0.012 +0.026}$ & $0.710_{-0.060 -0.113}^{+0.052 +0.122}$ & $14.32_{-0.19 -0.45}^{+0.33 +0.54}$ \\
~ & ~ & ~ & ~ & ~ & ~ & ~ & ~ \\

\hline
\end{tabular}
\end{center}
\caption{Summary of the results of the likelihood analysis of the models discussed in the text. The meaning of the entries is as follows. By writing $x = bf^{+\delta_{+} +\delta_{++}}_{-\delta_{-} -\delta_{ --}}$, we mean that $x$ is the maximum likelihood value of the considered quantity, while the $68\%$ and $95\%$ confidence ranges are $(x - \delta_{-}, x + \delta_{+})$ and  $(x - \delta_{--}, x + \delta_{++})$ respectively. We use this notation to give our constraints on the model parameters $(q_0, y, \Omega_M)$ and the derived quantities $({\cal{A}}, {\cal{R}}, z_T, t_0)$. For the $RK$ case, we do not estimate $({\cal{A}}, {\cal{R}}, z_T, t_0)$ since the model may be discarded. For the $MB$ and $Dt$ case, we are able to give only upper limits on $y$, while only lower limits may be given on this same quantity for the $PR$ model. See the text for discussion.}
\end{table*}

As a general result, we note that the constraints on both $q_0$ and $\Omega_M$ turn out to be essentially model independent. Moreover, they are consistent with recent estimates obtained using different datasets and dark energy models with a perfect fluid EoS (constant or redshift dependent). In particular, both values are quite similar to those predicted for the concordance $\Lambda$CDM model yielding $(q_0, \Omega_M) \simeq (-0.5, 0.3)$. Actually, this is not much surprising. As discussed in Sect.\,III, the four EoS considered mimic well the cosmological constant for small values of $z$. Therefore, we do expect similar values for $q_0$ and $\Omega_M$ since these quantities are both evaluated today when the difference among the concordance model and our ones may be hardly detected.

The only parameter directly characterizing the different models is therefore $y$ so that constraints on $y$ are indeed strongly model dependent. Beside, the physical range for this quantity must be set on a case by case basis thus obviously impacting the final estimate so that we discuss separately the results for each model.

\subsubsection{Redlich\,-\,Kwong}

As a first issue, it is important to assess what is the range explored for the parameter $y$. Looking at Eq.(\ref{eq: wrk}), it is clear that, in order to avoid unphysical divergences of the EoS, the condition $1 - \sqrt{3 - 2 \sqrt{2}} \alpha \eta(z) \ne 0$ must hold. Moreover, from Eq.(\ref{eq: brk}), we get the further constrain $y \ne 1/\sqrt{3 - 2 \sqrt{2}}$ in order $\beta$ to not be divergent. We have checked (analytically and numerically) that choosing $0 \le y \le 2$ ensures that all the conditions quoted above are fulfilled whatever is the redshift $z$. Performing the likelihood analysis discussed above, we have obtained the constraints reported in the first row of Table\,I. Quite surprisingly,  the range for $y$ is very narrow. Exploring the likelihood contours in the $3\,-\,D$ parameter space, we have found that there are indeed two {\it local} minima of the pseudo\,-\,$\chi^2$ defined above. Our procedure select the {\it absolute} minimum thus selecting a very small region. There is, however, also a more subtle motivation for the very stringent constraints obtained on $y$. Although not apparent from the upper left panel of Fig.\,\ref{fig: ez}, the energy density strongly depends on $y$ in the region $y \ge 1$ so that also small deviations from the best fit value leads to significant departures from the best fitting curve thus leading to very strong constraints on the model parameters.

Although the best fit curve reproduces very well the data, the $RK$ model may be excluded on the basis of physical considerations. Actually, for $(q_0, y, \Omega_M)$ in the parameter space individuated by the constraints reported, the energy density increases with $z$ faster than the matter one. As a result, the universe turns out to be ever accelerating (so that the estimated transition redshift is negative) and never undergoes a matter dominated epoch in the past. Even if we have not performed a detailed calculation, it is nevertheless clear that such a situation leads to severe problems with both structure formation and nucleosynthesis. Note that the same qualitative behaviour holds for the parameters taking values in the other local mimimum. Given these problems, we conclude that the $RK$ EoS may be discarded and will not be considered anymore in the following.

\subsubsection{Modified Berthelot}

Looking at Eqs.(\ref{eq: wmb}) and (\ref{eq: bmb}), it is clear that there are no physical motivations to impose an upper limit on $y$, the only constraints thus being $y \ge 0$. In the limit $y >> 1$, combining Eq.(\ref{eq: wmb}) with Eqs.(\ref{eq: defypar}) and (\ref{eq: bmb}), we get $w \simeq (\beta/\alpha)/\eta = (2 q_0 - 1)/[3 (1 - \Omega_M) \eta]$ so that the EoS does not depend on $y$ in this regime. As a consequence, the constraints on $y$ turns out to be quite weak and the likelihood analysis may put only upper limits. For the best fit values in the second row of Table\,I,  Eq.(\ref{eq: wmb}) reduces to the perfect fluid one with $w = (2 q_0 - 1)/[3 (1 - \Omega_M)] \simeq -1.12$. This result could suggest that the likelihood analysis argues in favor of no need of giving off the perfect fluid hypothesis. Actually, it is worth stressing that the marginalized likelihood function for $y$ is quite flat so that values of $y \ne 0$ are perfectly viable. Indeed, the $1 \sigma$ confidence range extends up to $y = 4.2$ thus showing that the $MB$ EoS provides a good match with the data even when it significantly differs from the simplest model $p = w \rho$ with $w$ a constant. 

Let us now briefly comment on the possibility to use the $MB$ EoS in the framework of UDE models. Should this approach be correct, the likelihood analysis should have returned $\Omega_M \simeq \Omega_b$, while such low values are excluded at more than $3 \sigma$ level. As such, one could conclude that the UDE approach may be rejected. Actually, one should still explore the possibility that Eq.(\ref{eq: wmb}) is an effective EoS and formally decompose the energy density $\rho_X$ as sum of $\rho_{dm}$ and $\rho_{de}$ with the first and second term referring to dark matter and dark energy respectively. The EoS of this dark energy term is then evaluated imposing $w_{MB} =  w_{de} \rho_{de}/(\rho_{dm} + \rho_{de})$. Imposing $\rho_{dm} = \Omega_{dm} \rho_{crit} (1 + z)^3$, one should set $\Omega_{dm} = \Omega_M - \Omega_b$ using the value of $\Omega_M$ determined above and a model independent estimate of $\Omega_b$. Investigating this scenario is outside our aim so that we do not speculate further on this interesting possibility.

It is interesting to discuss with some detail the constraints derived on some physical quantities coming from the likelihoods for the model parameters. As a first consistency check, we have estimated both the acoustic peak and the shift parameters ${\cal{A}}$ and ${\cal{R}}$. Even if we have explicitly introduced priors on them in the definition of the pseudo\,-\,$\chi^2$ in Eq.(\ref{eq: defchi}), it is nevertheless possible that the likelihood procedure selects a region of the parameter space giving values of ${\cal{A}}$ and ${\cal{R}}$ in disagreement with the imposed priors\footnote{Qualitatively, this could be understood noting that the main contribution to $\chi^2$ comes from the 157 SNeIa, while ${\cal{A}}$ and ${\cal{R}}$ gives only a modest contribution unless the model is unreasonably different from the best fit one.}. Actually, it turns out that ${\cal{A}}$ and ${\cal{R}}$ agree very well (within $1 \sigma$)  with the measured ones, although the maximum likelihood values are slightly larger than the estimated ones.

While $q_0< 0$, the universe has entered the epoch of accelerated expansion only for $z < z_T$, this latter being the transition redshift previously defined and constrained for the $MB$ model as reported in Table\,I. The maximum likelihood value $z_T = 0.692$ is more than $1.7 \sigma$ larger than the tentative model independent estimate of Riess et al. giving $z_T = 0.46 \pm 0.13$ \cite{Riess04}. It is worth noting, however, our value of $z_T$ is in good agreement with that predicted for the concordance $\Lambda$CDM model, being in this case $z_T = (2\Omega_{\Lambda}/\Omega_M)^{1/3} - 1 \simeq 0.671$. This is not much surprising given that, in the region of the parameter space selected, the $MB$ EoS mimics well that of the $\Lambda$ term over the redshift range probed by the data.

Finally, we consider the age of the universe obtaining $t_0 = 14.29 \ {\rm Gyr}$ as maximum likelihood value and the $2 \sigma$ confidence range extending from $13.88$ up to $14.72 \ {\rm Gyr}$. This result is in satisfactory agreement with previous model dependent estimates such as $t_0 = 13.24_{+0.41}^{+0.89} \ {\rm Gyr}$ from Tegmark et al. \cite{Teg03} and $t_0 = 13.6 {\pm} 0.19 \ {\rm Gyr}$ given by Seljak et al. \cite{Sel04}. Aging of globular clusters \cite{Krauss} and nucleochronology \cite{Cayrel} give model independent (but affected by larger errors) estimates of $t_0$ still in good agreement with our one. 

\subsubsection{Dieterici}

Setting the range of $y$ for the $Dt$ EoS is a subtle task. Imposing that $w$ never diverges and $\beta$ does not vanish leads to the conditions $y \ne 2 - [y/(1- \Omega_M)] \eta(z)$ and $y \ne 2$. There is, however, a further constraint motivated by numerical integrations of the continuity equation that turns out to become unstable for $y \ge 1$. In order to avoid this problem, we have searched for the constraints on $y$ in the region $0 \le y \le 1$ only. It comes out that the marginalized likelihood is quite flat so that the full range is well within $1 \sigma$ from the best fit value. Although this is quite disturbing from the point of view of constraining the model, this is encouraging since it shows that abandoning the perfect fluid EoS in favor of the $Dt$ one still gives a good match with the data. As a final remark, let us note that the acoustic peak and the shift parameters ${\cal{A}}$ and ${\cal{R}}$, the transition redshift $z_T$ and the age of the universe $t_0$ estimated for the $Dt$ model are in very good agreement with the same quantities obtained for the $MB$ case so that we refer the reader to what already said above.

\subsubsection{Peng\,-\,Robinson}

Eq.(\ref{eq: bpr}) shows that there are two values of $y$ such that the $PR$ EoS reduces to the perfect fluid one, namely $y = 0$ (so that $\alpha = 0$ and Eq.(\ref{eq: wmb}) reduces to $w_{PR} = \beta$) and $y = 1$ (giving $w_{PR} = 0$). We have checked that values of $y > 1$ give rise to models having some pathological behaviours in the past (for instance, unphysical divergence of $\eta(z)$ for high $z$) so that we have restricted our attention to the range $0 \le y \le 1$. Once again, the likelihood function is quite flat so that we are able only to give lower limits on $y$. It is noteworthy, however, that the best fit value is now $y = 0.9$, that is the $PR$ EoS does not reduce to that of the perfect gas. Finally, it is worth noting that the constraints on ${\cal{A}}$, ${\cal{R}}$, $z_T$ and $t_0$ obtained for the $PR$ model agree very well with those estimated for the $MB$ case so that we refer the reader to what already said.

\subsection{Degeneracy with the $\Lambda$CDM and QCDM models}

The results discussed above demonstrates that the $MB$, $Dt$ and $PR$ EoS give rise to cosmological models that are in good agreement with the considered dataset. On the other hand, also models with a dark energy having a perfect fluid EoS provides a very good match to the same dataset. This consideration suggests that a sort of degeneracy among the different EoS should exist. Investigating in detail this issue needs a detailed set of simulations in order to understand under which conditions such a degeneracy may be broken. Although this is outside our aim, we nevertheless provide a preliminary analysis that is sufficient to get an interesting feeling of the problem. To this aim, we implement a quite simple procedure. First, we select an EoS and set its characterizing parameters. Then, we generate a sample of SNeIa according to the theoretical luminosity distance for the model with the EoS chosen before. Note that the sample comprises the same number of SNeIa of the Riess et al. Gold sample \cite{Riess04} and have the same redshift and distance modulus error distribution. Finally, we fit to this dataset the $\Lambda$CDM ($p_X = - \rho_X$) and the QCDM ($p_X = w_X \rho_X$) model. In order to render our analysis as similar as possible to that in Riess et al., in the second case we impose the prior $\Omega_M = 0.27 \pm 0.04$ as done in \cite{Riess04}. The parameters used in the simulations and the constraints obtained on the $\Lambda$ and QCDM model parameters are summarized in Table\,II.

\begin{table}
\begin{center}
\begin{tabular}{|c|c|c|c|c|}
\hline
\multicolumn{2}{|c|}{Input Model} & $\Lambda$CDM & \multicolumn{2}{|c|}{QCDM} \\
\hline \hline
Id & $y, \Omega_M$ & $\Omega_M$ & $\Omega_M$ & $w_X$ \\
\hline  \hline
~ & ~ & ~ & ~ & ~ \\

$MB$ & $1.0, 0.28$ & $0.28_{-0.02 -0.04}^{+0.02 +0.04}$ & $0.30_{-0.04 -0.08}^{+0.04 +0.08}$ & $-1.06_{-0.16 -0.35}^{+0.12 +0.22}$ \\

~ & ~ & ~ & ~ & ~ \\

$Dt$ & $0.25, 0.28$ & $0.26_{-0.02 -0.04}^{+0.02 +0.04}$ & $0.34_{-0.04 -0.08}^{+0.03 +0.06}$ & $-1.27_{-0.17 -0.38}^{+0.15 +0.27}$ \\

~ & ~ & ~ & ~ & ~ \\

$PR$ & $0.35, 0.28$ & $0.27_{-0.02 -0.05}^{+0.03 +0.06}$ & $0.27_{-0.04 -0.08}^{+0.04 +0.08}$ & $-0.97_{-0.12 -0.26}^{+0.09 +0.18}$ \\

~ & ~ & ~ & ~ & ~ \\
\hline
\end{tabular}
\end{center}
\caption{Summary of the results of the likelihood analysis on the simulated dataset described in the text. The first two columns identifies the input model giving the EoS id and model parameters. In particular, for all the EoS, we have set $q_0 = -0.55$. The third column refers to the $\Lambda$CDM model, while fourth and fifth columns are for the QCDM model. Maximum likelihood values and confidence ranges are reported using the same scheme as in Table\,I.}
\end{table}

Some interesting lessons may be learned from this simple exercise. First, we note that the $\Lambda$CDM model fits well in all the cases considered. Moreover, the estimated $\Omega_M$ is quite close to the input value so that no systematic errors is induced on this parameter. Note that this result does not depend on the particular choice of the EoS parameters provided they lie in the confidence ranges summarized in Table\,I. Actually, such a result could be expected considering that, over the redshift range probed by the SNeIa sample and for the values of parameters chosen, the three EoS chosen mimic quite well the $\Lambda$CDM model so that it is not surprising that the simulated data may be fitted by the concordance scenario. It is still more interesting to consider the results from fitting the $QCDM$ model to the simulated dataset. While the estimated $\Omega_M$ is again quite similar to the input value (although biased and formally not in agreement within the errors), the constraints on $w_X$ may extend in the phantom region ($w_X < -1$) depending on the EoS adopted and the parameters chosen. This preliminary test suggests a possible way to escape the need of the problematic phantom dark energy (i.e. a negative pressure fluid with $w_X < -1$). Indeed, Table\,II shows that $w_X < -1$ may be the consequence of forcing the perfect fluid EoS to fit a cosmological model where the {\it true} dark energy EoS is not the perfect fluid one. This intriguing scenario have, however, to be further investigated with a more careful and extensive set of simulated dataset also taking into account other possible probes such as the priors on ${\cal{A}}$ and ${\cal{R}}$ or the gas mass fraction in galaxy clusters.

\section{The CMBR peaks position}

The analysis presented above has convincingly shown that dark energy models with EoS given by the $MB$, $Dt$ and $PR$ parametrizations are indeed viable alternatives with respect to the usual perfect fluid assumption. Indeed, the $r(z)$ diagram, the acoustic peak ${\cal{A}}$ and the shift parameter ${\cal{R}}$ are correctly predicted and also the estimated age of the universe is in good agreement with other estimates in literature. 

As a further test, we compute the positions of the first three peaks in the CMBR anisotropy spectrum using the procedure detailed in \cite{DorLil01,DorLil02}. According to this prescription, in a flat universe made out of a matter term and a scalar field\,-\,like fluid, the position of the $m$\,-\,th peak is given by\,:

\begin{equation}
l_m = l_A (m - \bar{\varphi} - \delta \varphi_m)
\label{eq: lm}
\end{equation}
with $l_A$ the acoustic scale, $\bar{\varphi}$ the overall peak shift and $\delta \varphi_m$ the relative shift of the $m$\,-\,th peak with respect to the first. While $\bar{\varphi}$ and $\delta \varphi_m$ are given by the approximated formulae in Ref.\,\cite{DorLil02}, the acoustic scale for flat universes may be evaluated as \cite{DorLil01}\,:

\begin{eqnarray}
l_A & = & \frac{\pi}{\bar{c}_s} \left \{ \frac{F}{\sqrt{1 - \bar{\Omega}_{ls}^{\phi}}} \right . \nonumber \\
~ & \times & \left . \left [ 
\sqrt{a_{ls} + \frac{\Omega_0^r}{1 - \Omega_0^{\phi}}} - \sqrt{\frac{\Omega_0^r}{1 - \Omega_0^{\phi}}} \right ]^{-1} - 1 \right \}
\label{eq: la}
\end{eqnarray}
with\,:

\begin{equation}
F = \frac{1}{2} \int_{0}^{1}
{da \left [ a + \frac{\Omega_0^{\phi} a^{1 - 3\bar{w}_0} + \Omega_0^r (1 - a)}{1 - \Omega_0^{\phi}} \right ]^{-1/2}}
\label{eq: defeffe}
\end{equation}
where we use Eq.(\ref{eq: zls}) to determine $a_{ls} = (1 + z_{ls})^{-1}$. The other quantities entering Eqs.(\ref{eq: la}) and (\ref{eq: defeffe}) are defined as follows \cite{DorLil01,DorLil02}\,:

\begin{figure*}
\centering \resizebox{17cm}{!}{\includegraphics{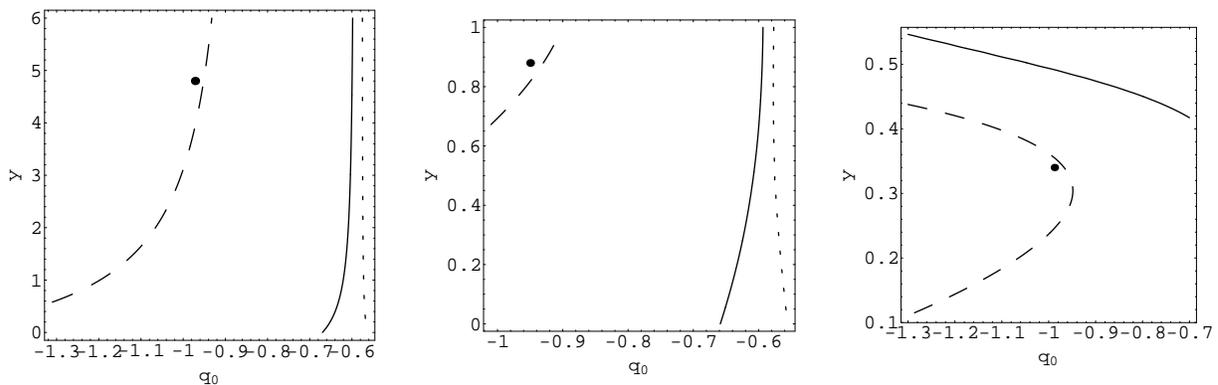}}
\caption{Constraints on the deceleration parameter $q_0$ and the scaled density parameter $y$ for the $MB$ (left panel), $Dt$ (central panel) and $PR$ (right panel) EoS. Models with parameters on the right of the short, solid and loong dashed lines give values of $(l_1, l_2, l_3)$ respectively in agreement within $1 \sigma$ with the measured ones. The black dot individuates the best fit model discussed in the text. For all the EoS, $\Omega_M$ is set to the best fit value reported in Table\,I. }
\label{fig: peaks}
\end{figure*}

\begin{equation}
\bar{c}_s = \frac{1}{\tau_{ls}} \int_{0}^{\tau_{ls}}{\left [ 3 + \frac{9}{4} \frac{\rho_b(\tau)}{\rho_r(\tau)} \right ]^{-2} d\tau} \ ,
\label{eq: defcs}
\end{equation}

\begin{equation}
\bar{w}_0 = \frac{\int_0^{\tau_0}{\Omega_{\phi}(\tau) w(\tau) d\tau}}{\int_0^{\tau_0}{\Omega_{\phi}(\tau) d\tau}} \ ,
\label{eq: wzmean}
\end{equation}

\begin{equation}
\bar{\Omega}_{ls}^{\phi} = \frac{1}{\tau_{ls}} \int_{0}^{\tau_{ls}}{\Omega_{\phi}(\tau) d\tau} \ ,
\label{eq: meanom}
\end{equation}
where $\tau = \int{a^{-1} dt}$ is the conformal time, $\rho_b$ and $\rho_r$ are the energy densities of the baryons and radiation respectively, $w(z)$ and $\Omega_{\phi} = \rho_{\phi}/\rho_{crit}(z)$ are the barotropic factor and the density parameter of the scalar field. In order to use Eqs.(\ref{eq: la})\,-\,(\ref{eq: meanom}), we note that the role of the scalar field fluid is played by the dark energy so that all the quantities with the subscript $\phi$ have now to be evaluated using the energy density corresponding to a given EoS. Finally, we set the present day value of the radiation density parameter as $\Omega_0^r = 9.89 \times 10^{-5}$ \cite{DorLil01,DorLil02} and $n = 1$ as index of the spectrum of primordial fluctuations, entering the approximated formulae for $\bar{\varphi}$ and $\delta \varphi_m$.

The position of the first two peaks in the CMBR anisotropy spectrum ha been determined with great accuracy by WMAP giving \cite{WMAP}\,:

\begin{equation}
l_1^{WMAP} = 220.1 \pm 0.08 \ \ , \ \ l_2^{WMAP} = 546 \pm 10 \ \ ,
\label{eq: l12obs}
\end{equation}
while the position of the third peak is more uncertain and may be estimated as \cite{Boom}\,:

\begin{equation}
l_3^{Boom} = 851 \pm 31 \ .
\label{eq: l3obs}
\end{equation}
Having only three data points, it is clear that only qualitative constraints can be imposed on the model parameters. For this reason, we fix $\Omega_M$ to the best fit value in Table\,I for each EoS and use a $\chi^2$ analysis to constrain $q_0$ and $y$. Formally, the best fit parameters turn out to be\,:

\begin{equation}
(q_0, y) = \left \{
\begin{array}{ll}
(-0.972, 4.80) & {\rm for \ the \ } MB \ {\rm EoS} \ , \\
(-0.950, 0.88) & {\rm for \ the \ } Dt \ {\rm EoS} \ , \\
(-0.988, 0.34) & {\rm for \ the \ } PR \ {\rm EoS} \ , \\
\end{array}
\right .
\label{eq: bfpeaks}
\end{equation}  
giving $(l_1, l_2, l_3) = (208.3, 546.3, 857.2)$ as best fit values independent of the EoS considered. However, as Fig.\,\ref{fig: peaks} shows, the region of the parameter space $(q_0, y)$ that are consistent within $1 \sigma$ with the bounds from the position of the peaks is quite large, so that, as already predicted, only weak constraints can be derived. Although a detailed fit to the full CMBR anisotropy spectrum is needed, this preliminary analysis gives encouraging results. Indeed, considering the most stringent cut (that on $l_1$), Fig.\,\ref{fig: peaks} shows that it is possible to find out models that are in agreement with both the fit to the dimensionless coordinate distances and the position of the first three peaks. 

\section{Scalar field potential}

Although the agreement with the observations is a valid motivation for these models, it is nonetheless important to look for a theoretical approach to further substantiate our proposal. Such a scheme may be easily recovered in the framework of scalar field quintessence. In such a case, the energy density and the pressure of the dark energy fluid read\,:

\begin{equation}
\rho_{\phi} = \frac{1}{2} \dot{\phi}^2 + V(\phi) \ ,
\label{eq: rhophi}
\end{equation}

\begin{equation}
p_{\phi} = \frac{1}{2} \dot{\phi}^2 - V(\phi) \ ,
\label{eq: pphi}
\end{equation}
where $\phi$ is the scalar field evolving under the action of the self\,-\,interaction potential $V(\phi)$. For a given $V(\phi)$, Eqs.(\ref{eq: rhophi}) and {\ref{eq: pphi}) may be inserted in the Friedmann equations in order to determine $\rho_{\phi}(z)$, $w_{\phi}(z) = p_{\phi}(z)/\rho_{\phi}(z)$ and the other dynamical quantities of interest. It is worth noting, however, that this procedure may be reverted so that, for given $w(z)$ and $E(z)$, one may find out the self\,-\,interaction potential $V(\phi)$ giving rise to that kind of cosmological expansion. To this end, one may first determine $d\phi(z)/dz$ and $V(z)$ as \cite{Guo} (but see also \cite{rec} for reconstruction from the SNeIa data directly)\,:

\begin{equation}
\tilde{V}(z) = \frac{1}{2}(1 - w_{\phi}) E(z) \ ,
\label{eq: vz}
\end{equation}

\begin{equation}
\frac{d \tilde{\phi}(z)}{dz} = - \frac{\sqrt{3 (1 + w_{\phi})}}{1 + z} \left [ 1 + \frac{\Omega_M (1 + z)^2}{(1 - \Omega_M) E(z)} \right ]^{-1/2}
\label{eq: dphidz}
\end{equation}
where we have defined $\tilde{V} \equiv V/\rho_{\phi}(z = 0)$ and $\tilde{\phi} = \phi/M_{Pl}$ with $\rho_{\phi}(0) = (1 - \Omega_M) \rho_{crit}$ and $M_{Pl} = (8 \pi G)^{-1/2}$. Note that, to get Eq.(\ref{eq: dphidz}), we have chosen $\dot{\phi} > 0$ (which gives $d\phi/dz < 0$) without any loss of generality. Eq.(\ref{eq: vz}) gives $V(z)$, while $V(\phi)$ may be obtained integrating Eq.(\ref{eq: dphidz}) with the initial condition $\phi(z = 0) = 0$ to get $\phi(z)$ and then inverting this relation with respect to $z$.

\begin{figure}
\centering \resizebox{8.5cm}{!}{\includegraphics{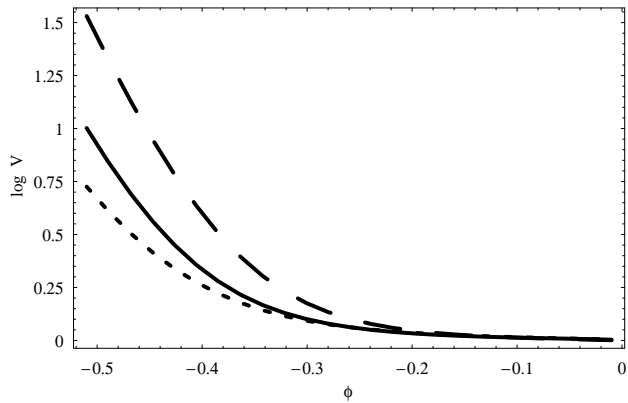}}
\caption{Reconstructed scalar field potential over the redshift range $(0, 10)$ for the models with the $MB$ (short dashed), $Dt$ (solid) and $PR$ (long dashed) EoS. The potential is normalized to be 1 at $z = 0$, while, on the abscissa, $\phi$ is in units of the Planck mass $M_{Pl}$.}
\label{fig: vpot}
\end{figure}

We have applied this procedure imposing $w_{\phi}(z) = w_i(z)$ (with $i = MB, Dt, PR$) over the redshift range $z = (0, 10)$ in order to recover the scalar field potential giving rise to our exotic EoS. Not surprisingly, an analytical solution is not possible so that we have resorted to numerical techniques setting the model parameters to their best fit values. The results are shown in Fig.\,\ref{fig: vpot} for the $MB$, $Dt$ and $PR$ EoS. It is worth noting that, although the three EoS are quite different, the potential $V(\phi)$ is remarkably similar and any difference could be hardly detected over a large range in $\phi$. As a matter fact, the same analytical approximating function may be fitted to the three models. Indeed, we find that, within $2\%$, a very good fitting is obtained for\,:

\begin{equation}
V(\phi) = V_0 \left \{ 1 + V_s \left ( \frac{\phi}{\phi_s} \right )^{\alpha} 
\exp{\left [ \frac{1}{2} \left ( \frac{\phi}{\phi_s} \right )^2 \right ]} \right \}
\label{eq: vfit}
\end{equation}
with $(V_s, \phi_s, \alpha)$ fitting parameters to be determined on a case by case basis. For the best fit models, we get\,:

\begin{equation}
(V_s, \phi_s, \alpha) = \left \{
\begin{array}{l}
(0.0448, -0.1817, 0.6098) \\
(0.0331, -0.1611, 0.5197) \\
(0.0369, -0.1517, 0.9434) \\
\end{array}
\right . 
\nonumber
\end{equation}
for the $MB$, $Dt$ and $PR$ EoS respectively. Considering the behaviour of $w(z)$ at high $z$ shown in Fig.\,\ref{fig: wz}, it is somewhat surprising that the same functional expression approximates well the potential $V(\phi)$ for all the three EoS. However, we have checked that there are no systematic errors in the reconstruction procedure. Indeed, for the $Dt$ and $PR$ EoS, at high enough $z$, $\dot{\phi}^2$ is negligible with respect to $V(\phi)$ so that the scalar field enters in a slow roll like regime and $w \simeq -1$ as in the lower panels of Fig.\,\ref{fig: wz}. For the $MB$ case, a slow roll regime is not achieved at high $z$ where, on the contrary, $\dot{\phi^2} \simeq V(\phi)$ and hence the EoS of the scalar field counterpart vanishes. 

It is worth noting that the approximating potential is quite different from those often used in literature, such as the exponential potential \cite{exp,RS} and the power law one \cite{RP}. On the other hand, its shape is the same as those proposed in supergravity inspired models according to which it is $V \propto \phi^{\alpha} \exp{(\phi^2)}$ \cite{SUGRA}. However, $\alpha \le 0$ in such models, while we find $\alpha$ positive. While for large $\phi/\phi_s$ both our approximating potential and SUGRA\,-\,like ones are exponential, for small $\phi/\phi_s$, $V(\phi)$ takes a power law shape in the SUGRA case, while, for our models, $V(\phi)$ is approximately constant so that a cosmological constant behaviour is achieved. This is consistent with the results $w(z = 0) \simeq -1$ we find for the present value of the EoS using the best fit parameters in Table\,I.

\section{Conclusions}

Notwithstanding their (somewhat radically) different approaches to the dark energy puzzle, all the models proposed so far assumes that the dark energy EoS is a linear function of the energy density. From a thermodynamical point of view, the ansatz $p = w \rho$ (no matter whether $w$ is a constant or a function of $z$) means that the fluid is modeled as a perfect gas. Putting forward the analogy with classical thermodynamics, however, it is well known that the perfect fluid approximation is quite crude and it is, in particular, unable to deal with critical phenomena such as phase transitions and the behaviour of the fluid near critical points. As a matter of fact, the perfect fluid approximation works only in very particular conditions. On the other hand, our deep ignorance of the fundamental properties of the dark energy nature does not motivate the choice of such idealized condition for the present state of this component. It is thus interesting to consider EoS that are more general than the perfect fluid one reducing to this latter in a certain regime. 

Motivated by these considerations, we have investigated here the consequences of abandoning the perfect fluid approximation on the dynamics of a dark energy dominated universe. To this end, we have considered four different EoS, namely the Redlich\,-\,Kwong Eq.(\ref{eq: wrk}), the Modified Berthelot Eq.(\ref{eq: wmb}), the Dieterici Eq.(\ref{eq: wdt}) and the Peng\,-\,Robinson Eq.(\ref{eq: wpr}) parametrizations. These have been chosen because, from classical thermodynamics, we know they are well behaved also in critical conditions. The viability of the models and constraints on their characterizing parameters have been studied by using a likelihood analysis taking into account the observations on the dimensionless coordinate distance to SNeIa and radiogalaxies and priors on the acoustic peak and shift parameters ${\cal{A}}$ and ${\cal{R}}$. This test has shown that all the four EoS are able to give rise to models that fits quite well the available dataset, but the $RK$ EoS has to be rejected since it does not give rise to a deceleration phase in the past. On the other hand, the $MB$, $Dt$ and $PR$ EoS predict reasonable values for the transition redshift $z_T$ and the age of the universe $t_0$ in good agreement with previous model independent estimates. As a further check, we have also evaluated the position of the first three peaks in the CMBR anisotropy spectrum finding out that, for each model, there exists a region of the parameter space such that both the CMBR peaks and the $r(z)$ diagram are correctly reproduced. These successful results are quite encouraging since they show that the perfect fluid EoS may be given off without worsening the agreement with the data. This nice consideration strongly motivates further test of these models in order to both better constrain their parameters and try to select among them according to what model is better suited to describe what is observed. Since the $MB$, $Dt$ and $PR$ EoS evolve with $z$ in different ways, it is desirable to resort to observables depending on $w(z)$ rather than its value over only a limited redshift range. Interesting candidates, from this point of view, are the full CMBR spectrum (not only its peaks position) and the growth factor that also depends on the theory of perturbations. 

As an interesting byproduct of the likelihood test, we have discovered a degeneracy with both the concordance $\Lambda$CDM and the quiessence (dark energy with constant $w$) QCDM models. Actually, we have fitted both $\Lambda$CDM and QCDM models to Gold\,-\,like SNeIa dataset simulated using one of our EoS as true background cosmological model. Using the same procedure in Riess et al. \cite{Riess04}, we have found that the $\Lambda$CDM model provides a very good fit for values of $\Omega_M$ in well agreement with the input ones. On the contrary, the QCDM model still gives a good match with the simulated data, but $\Omega_M$ is slightly biased high and $w$ may be artificially pushed in the phantom region $w < -1$. This suggests the intriguing possibility that phantom models turns out to be the best fit to SNeIa data only because of a systematic error on the EoS. It is worth stressing, however, that this result is only preliminary being based on a limited dataset. To further substantiate it, one should carry out a Fisher matrix analysis or detailed Montecarlo simulations also taking into account other probes as the full CMBR anisotropy spectrum.

Having been inspired by classical thermodynamics, the EoS considered are phenomenologically motivated, but lacks a background theoretical model. To overcome this difficulty, we have worked out a scalar field interpretation reconstructing the self\,-\,interaction potential in such a way that the quintessence EoS is the same as the $MB$, $Dt$ or $PR$ EoS. The potential $V(\phi)$ thus obtained is very well approximated by an analytical expression that is neither exponential\,-\,like nor power law\,-\,like, but has formally the same form as the SUGRA inspired models. However, there is a significant difference since, for small $\phi$, both potentials scale as $\phi^{\alpha}$, but $\alpha > 0$ for our models rather than negative as in the SUGRA scenario. It could be interesting to work out the consequences of such a difference. On the other hand, it is also possible that the EoS considered have to be considered as {\it effective} ones as is the case in some braneworld inspired dark energy models \cite{DGP} or for the curvature fluid \cite{capozcurv,review,noicurv} in $f(R)$ theories.

We would like to conclude with a general comment. As it is well known, the perfect fluid EoS is only a crude approximation of a {\it real} fluid that is usually used in cosmology since it represents the simplest way to fit the available data. However, as first shown in the Van der Waals quintessence scenario \cite{vdw,Kr03} and further demonstrated here, abandoning the perfect fluid EoS still makes it possible to fit the available astrophysical data with the same accuracy so that the use of realistic EoS turns out to be motivated also {\it a posteriori}. In our opinion, the era of {\it precision cosmology} calls for {\it precision theory} so that time is come to abandon approximate description such as the perfect fluid one. Which is the {\it most realistic} description of the dark energy term is a topic that worths be addressed with the help of hints coming from thermodynamic analogies. 

\appendix

\section{Some details on the EoS}

Although inspired by classical thermodynamics, the EoS we have considered are somewhat exotic so that we believe it is useful to give some further details from the thermodynamic point of view. As a preliminary step, let us denote with $(p, V, T)$ the pressure, the volume and the temperature of the fluid and with a subscript $c$ these quantities evaluated at the {\it critical point}. Let us also remember that the critical point is defined by the conditions\,:

\begin{equation}
\left ( \frac{\partial p}{\partial V} \right )_{T = cst} = 
\left ( \frac{\partial^2 p}{\partial V^2} \right )_{T = cst} = 0 \ .
\label{eq: crit}
\end{equation}
Let us also denote with the subscript $r$ the reduced quantities, i.e. $x_r \equiv x/x_c$. In what follows, we show how Eqs.(\ref{eq: wrk})\,-\,({\ref{eq: wpr}) are obtained starting from their thermodynamical analog. As a general remark, note that the temperature of the dark energy fluid should be intended as an {\it effective} rather than a {\it physical} one in order to avoid problems with negative values. Note that this problem is also present in the case of the perfect fluid EoS where a negative $w < 0$ is formally equivalent to a negative $T$.

\subsection{Redlich\,-\,Kwong}

Let us first consider the $RK$ EoS that is defined as\,:

\begin{equation}
p = \frac{R T}{V - b} - \frac{a}{V (V + b) T^{1/2}}
\label{eq: prk}
\end{equation}
with $R$ the gas constant and $(a, b)$ two model parameters. Inserting Eq.(\ref{eq: prk}) into Eq.(\ref{eq: crit}), we get\,:

\begin{displaymath}
a = \sqrt{3 - 2 \sqrt{2}} \ R T_c^{3/2} V_c \ \ \ \ , \ \ \ \ 
b = (1 - \sqrt{2}) V_c 
\end{displaymath}
so that Eq.(\ref{eq: prk}) may be rewritten as\,:

\begin{equation}
p = \tilde{p_c} \times \frac{(T_r V_r^2)^{-1/2}}{(1 - \sqrt{2})^{-1} V_r - 1} 
\left ( \frac{T_r^{3/2} V_r}{\sqrt{3 - 2 \sqrt{2}}} - 1 \right )
\label{eq: prkbis}
\end{equation}
with\,:

\begin{displaymath}
\tilde{p_c} = \frac{\sqrt{2 (3 - 2 \sqrt{2})}}{(1 - \sqrt{2})(1 - \sqrt{3 - 2 \sqrt{2}})} \times p_c \ , 
\end{displaymath}
\begin{displaymath}
p_c = \frac{1 - \sqrt{3 - 2 \sqrt{2}}}{\sqrt{2}} \times \frac{R T_c}{V_c} 
\simeq 0.414 \ R T_c/V_c \ .
\end{displaymath} 
Note that the critical pressure is almost half the perfect fluid one. Let us now assume that the temperature is constant and equal to its critical value\footnote{The ansatz $T = T_c$ is somewhat arbitrary, but does not lead to any loss of generality. Actually, choosing a different $T$ only rescales the EoS. Since we do not know either $T$ or $T_c$, it is a useful working hypothesis to set $T_r = 1$.} so that $T_r = 1$. Using then $V = 1/\rho \rightarrow V_r = \rho_c/\rho$, after some algebra, Eq.(\ref{eq: prkbis}) may be finally rewritten as\,:

\begin{displaymath}
\tilde{p} = \beta \eta \times 
\frac{1 - \sqrt{3 - 2 \sqrt{2}} \alpha \eta}{1 - (1 - \sqrt{2}) \alpha \eta} 
\end{displaymath}
which is the same as Eq.(\ref{eq: wrk}) having posed $\tilde{p} = p/\rho_{crit}$, $\eta = \rho/\rho_{crit}$ and defined\,:

\begin{displaymath}
\alpha = \rho_{crit}/\rho_c \ \ \ \ , \ \ \ \
\beta = \frac{\sqrt{2}}{1 - \sqrt{3 - 2 \sqrt{2}}} \times p_c/\rho_{c} \ .
\end{displaymath}
Note that, for the best fit parameters in Table\,1, $\beta < 0$ so that $p_c < 0$ as expected since dark energy is known to have negative pressure.

\subsection{Modified Berthelot}

Let us summarize here the main steps above for the case of the $MB$ EoS. The pressure $p$ as function of temperature $T$ and volume $V$ is implicitly defined as\,:

\begin{equation}
p = \frac{R T}{V} \left [ 1 + \frac{9}{128} \left ( \frac{p}{p_c} \right )
\left ( \frac{T}{T_c} \right )^{-1}  \left ( 1 - \frac{T_c^2}{T^2} \right ) \right ]
\label{eq: pmb}
\end{equation}
where the critical pressure is given as\,:

\begin{displaymath}
p_c = \frac{83}{128} \frac{R T_c}{V_c} \ .
\end{displaymath}
Using this relation, we may rewrite Eq.(\ref{eq: pmb}) as\,:

\begin{equation}
p = \frac{128}{83} p_c \times \frac{T_r/V_r}{1 - (9/128) V_r^{-1} (1 - 6 T_r^{-2})} \ .
\label{eq: pmbbis}
\end{equation}
Setting $T_r = 1$ and $V_r = \rho_c/\rho$, after some simple algebra, we finally get\,:

\begin{displaymath}
\tilde{p} = \frac{\beta \eta}{1 + \alpha \eta}
\end{displaymath}
with $\tilde{p} = p/\rho_{crit}$, $\eta = \rho/\rho_{crit}$ and we have defined\,:

\begin{displaymath}
\alpha = \frac{45 \rho_{crit}}{128 \rho_c} \ \ \ \ , \ \ \ \ 
\beta = \frac{128^2}{45 \times 83} \frac{p_c}{\rho_c} \simeq 4.4 (p_c/\rho_c) \ .
\end{displaymath}
As for the $RK$ case, the best fit parameters in Table\,1 gives $\beta < 0$ as expected.

\subsection{Dieterici}

In terms of thermodynamical quantities, the $Dt$ EoS reads\,:

\begin{equation}
p = \frac{R T}{V - b} \exp{\left ( - \frac{a}{R T V} \right )} 
\label{eq: pdt}
\end{equation}
where the two parameters $(a, b)$ may be determined solving Eq.(\ref{eq: crit}) thus obtaining\,:
\begin{displaymath}
a = 2 R T_c V_c \ \ \ \ , \ \ \ \ b = V_c/2 \ .
\end{displaymath}
Introducing reduced variables $(T_r, V_r)$ and using the expression for the critical pressure\,:

\begin{displaymath}
p_c = \frac{2 R T_c}{e^2 V_c} \ ,
\end{displaymath}
we rewrite Eq.(\ref{eq: pdt}) as\,:

\begin{equation}
p = \frac{p_c T_r}{2 V_r - 1} \exp{\left [ 2 (1 - T_r^{-1} V_r^{-1} \right ]} \ .
\label{eq: pdtbis}
\end{equation}
With the usual positions $T_r = 1$ and $V_r = \rho_c\rho$, the above relation finally becomes\,:

\begin{displaymath}
\tilde{p} = \frac{\beta \eta}{2 - \alpha \eta} \exp{\left [ 2 (1 - \alpha \eta) \right ]}
\end{displaymath}
which is the same as Eq.(\ref{eq: wdt}) provided one defines tilted quantities as usual and

\begin{displaymath}
\alpha = \rho_{crit}/\rho \ \ \ \ , \ \ \ \  \beta = p_c/\rho_c \ .
\end{displaymath}
Not surprisingly, $\beta$ (and hence $p_c$) turns out to be negative for the best fit parameters.

\subsection{Peng\,-\,Robinson}

As a final case, let us consider the $PR$ EoS starting from its expression in terms of thermodynamical quantities\,:

\begin{equation}
p = \frac{R T}{V - b} - \frac{a}{V (V + b) + b (V - b)} \ .
\label{eq: ppr}
\end{equation}
Solving the equations for the critical points allows us to express $(a, b)$ in terms of $(T_c, V_c)$ giving\,:

\begin{displaymath}
a = c_a R T_c V_c \ \ \ \ , \ \ \ \ b = c_b V_c
\end{displaymath}
with $c_a \simeq 1.487$ and $c_b \simeq 0.253$. The critical pressure turns out to be \,:

\begin{displaymath}
p_c = \frac{R T_c/V_c}{1 - c_b} \left [ 1 - \frac{c_a}{(1 + c_b)/(1 - c_b) + c_b} \right ] 
\simeq 0.307 R T_c/V_c \ .
\end{displaymath}
Introducing reduced variables, Eq.(\ref{eq: ppr}) rewrites as\,:

\begin{equation}
p = \frac{p_c T_r c_p^{-1}}{V_r - c_b} \left [ 1 - \frac{c_a T_r^{-1}}
{V_r (V_r + c_b)/(V_r - c_b) + c_b} \right ] \ .
\label{eq: pprbis}
\end{equation}
Eq.(\ref{eq: wpr}) is finally obtained by setting in the above relation $T_r = 1$ and $V_r = \rho_c/\rho$ thus getting\,:

\begin{displaymath}
\tilde{p} = \frac{\beta \eta}{1 - \alpha \eta} \left [ 1 - \frac{(c_a/c_b) \alpha \eta}
{(1 + \alpha \eta)/(1 - \alpha \eta) + \alpha \eta} \right ]
\end{displaymath}
with $\tilde{p}$ and $\eta$ defined as usual, while we have set\,:

\begin{displaymath}
\alpha = \rho_{crit}/\rho_c \ \ \ \ , \ \ \ \ 
\beta = \frac{p_c/\rho_c}{c_b c_p} \ .
\end{displaymath}
Once again, the best fit parameters gives $\beta < 0$ and hence $p_c < 0$ as expected.


\begin{thebibliography}{99}

\bibitem{Boom}
P. de Bernardis et al. 2000, Nature, 404, 955

\bibitem{CMBR}
R. Stompor et al., ApJ, 561, L7, 2001; 
C.B. Netterfield et al., ApJ, 571, 604, 2002; 
R. Rebolo et al., MNRAS, 353, 747, 2004

\bibitem{WMAP}
D.N. Spergel et al., ApJS, 148, 175, 2003

\bibitem{SNeIa}
A. G. Riess et al., AJ, 116, 1009, 1998; 
S. Perlmutter et al., ApJ, 517, 565, 1999; 
R.A. Knop et al., ApJ, 598, 102, 2003; 
J.L. Tonry et al., ApJ, 594, 1, 2003; 
B.J. Barris et al., ApJ, 602, 571, 2004; 

\bibitem{Riess04}
A.G. Riess et al., ApJ, 607, 665, 2004

\bibitem{LSS}
S. Dodelson et al., ApJ, 572, 140, 2002; 
W.J. Percival et al., MNRAS, 337, 1068, 2002; 
A.S. Szalay et al., ApJ, 591, 1, 2003; 
E. Hawkins et al., MNRAS, 346, 78, 2003; 
A.C. Pope et al., ApJ, 607, 655, 2004

\bibitem{Lyalpha}
R.A.C. Croft et al., ApJ, 495, 44, 1998; 
P. McDonald et al., astro\,-\,ph/0405013, 2004

\bibitem{Lambda}
S.M. Carroll, W.H. Press, E.L. Turner, ARAA, 30, 499, 1992;
V. Sahni, A. Starobinski, Int. J. Mod. Phys. D, 9, 373, 2000

\bibitem{Teg03}
M. Tegmark et al., Phys. Rev. D, 69, 103501, 2004 

\bibitem{Sel04}
U. Seljak et al., Phys. Rev. D, 71, 103515, 2005

\bibitem{QuintRev}
P.J.E. Peebles, B. Rathra, Rev. Mod. Phys., 75, 559, 2003;
T. Padmanabhan, Phys. Rept., 380, 235, 2003

\bibitem{Chaplygin}
A. Kamenshchik, U. Moschella, V. Pasquier, Phys. Lett. B, 511, 265, 2001;
N. Bili\'c, G.B. Tupper, R.D. Viollier, Phys. Lett. B, 535, 17, 2002;
M.C. Bento, O. Bertolami, A.A. Sen, Phys. Rev. D, 67, 063003

\bibitem{tachyon}
G.W. Gibbons, Phys. Lett. B, 537, 1, 2002;
T. Padmanabhan, Phys. Rev. D, 66, 021301, 2002;
T. Padmanabhan, T.R. Choudury, Phys. Rev. D, 66, 081301, 2002;
J.S. Bagla, H.K. Jassal, T. Padmanabhan, Phys. Rev. D, 67, 063504, 2003;
E. Elizalde, S. Nojiri, S.D.Odintsov, hep\,-\,th/0405034, 2004

\bibitem{Hobbit}
V.F. Cardone, A. Troisi, S. Capozziello, Phys. Rev. D, 69, 083517, 2004;
S. Capozziello, A. Melchiorri, A. Schirone, Phys. Rev. D, 70, 101301, 2004

\bibitem{ie}
V.F. Cardone, A. Troisi, S. Capozziello, astro\,-\,ph/0506371, Phys. Rev. D accepted

\bibitem{Od}
S. Nojiri, S.D. Odintsov, hep\,-\,th/0505215, 2005

\bibitem{Odphan}
S. Npojiri, S.D. Odintsov, hep\,-\,th/0506212, 2005

\bibitem{DGP}
G.R. Dvali, G. Gabadadze, M. Porrati, Phys. Lett. B, 485, 208, 2000;
G.R. Dvali, G. Gabadadze, M. Kolanovic, F. Nitti, Phys. Rev. D, 64, 084004, 2001;
G.R. Dvali, G. Gabadadze, M. Kolanovic, F. Nitti, Phys. Rev. D, 64, 024031, 2002;
A. Lue, R. Scoccimarro, G. Starkman, Phys. Rev. D, 69, 124015, 2004

\bibitem{capozcurv}
S. Capozziello, Int. J. Mod. Phys. D, 11, 483, 2002

\bibitem{review}
S. Capozziello, S. Carloni, A. Troisi, Recent Research Developments in Astronomy and Astrophysics, Research Signpost Publisher, astro\,-\,ph/0303041, 2003

\bibitem{noicurv}
S. Capozziello, V.F. Cardone, S. Carloni, A. Troisi, Int. J. Mod. Phys. D, 12, 1969, 2003; 
S. Capozziello, V.F. Cardone, M. Francaviglia, astro\,-\,ph/0410135;
S. Carloni, P.K.S. Dunsby, S. Capozziello, A. Troisi, gr\,-\,qc/0410046, 2004

\bibitem{MetricRn}
S. Nojiri, S.D. Odintsov, Phys. Lett. B, 576, 5, 2003; 
S. Nojiri, S.D. Odintsov, Mod. Phys. Lett. A, 19, 627, 2003; 
S. Nojiri, S.D. Odintsov, Phys. Rev. D, 68, 12352, 2003; 
S.M. Carroll, V. Duvvuri, M. Trodden, M. Turner, Phys. Rev. D, 70, 043528, 2004

\bibitem{CCT}
S. Capozziello, V.F. Cardone, A. Trosi, Phys. Rev. D, 71, 043503, 2005

\bibitem{PalRn}
D.N. Vollick, Phys. Rev. 68, 063510, 2003; 
X.H. Meng, P. Wang, Class. Quant. Grav., 20, 4949, 2003; 
E.E. Flanagan, Phys. Rev. Lett. 92, 071101, 2004; 
E.E. Flanagan, Class. Quant. Grav., 21, 417, 2004; 
X.H. Meng, P. Wang, Class. Quant. Grav., 21, 951, 2004; 
G.M. Kremer and D.S.M. Alves, Phys. Rev. D, 70, 023503, 2004

\bibitem{lnR}
S. Nojiri, S.D. Odintsov, Gen. Rel. Grav., 36, 1765, 2004; 
X.H. Meng, P. Wang, Phys. Lett. B, 584, 1, 2004

\bibitem{Allemandi}
G. Allemandi, A. Borowiec, M. Francaviglia, Phys. Rev. D, 70, 043524, 2004; 
G. Allemandi, A. Borowiec, M. Francaviglia, Phys. Rev. D, 70, 103503, 2004; 
G. Allemandi, A. Borowiec, M. Francaviglia, S.D. Odintsov, gr\,-\,qc/0504057, 2005

\bibitem{nonparam}
U. Alam, V. Sahni, D. Saini, A.A. Starobinsky, MNRAS, 354, 275, 2004;
H.K. Jassal, J.S. Bagla, T. Padmanabhan, MNRAS, 356, 11, 2005

\bibitem{Rowlinson}
Rowlinson J.S., Widom B. 1982, {\it Molecular Theory of Capillarity}, Oxford University Press, Oxford (UK)

\bibitem{vdw}
S. Capozziello, S. De Martino, M. Falanga, Phys. Lett. A, 299, 494, 2002;
S. Capozziello, V.F. Cardone, S. Carloni, S. De Martino, M. Falanga, A. Troisi, M. Bruni, JCAP, 0504, 005, 2005

\bibitem{Kr03}
Kremer G.M. 2003, Phys. Rev. D, 68, 123507;
Kremer G.M. 2004, Gen. Rel. Grav., 36, 1423 

\bibitem{CosmoBooks}
Peebles P.J.E. 1993, {\it Principle of physical cosmology}, Princeton Univ. Press, Princeton (USA);
Peacock J. 1999, {\it Cosmological physics}, Cambridge University Press, Cambridge (UK)

\bibitem{DD04}
R.A. Daly, S.G. Djorgovski, ApJ, 612, 652, 2004

\bibitem{RGdata}
Guerra E.J., Daly R.A., Wan L. 2000, ApJ, 544, 659;
Daly R.A., Guerra E.J. 2002, AJ, 124, 1831;
Podariu S., Daly R.A., Mory M.P., Ratra B. 2003, ApJ, 584, 577;
Daly R.A., Djorgovski S.G. 2003, ApJ, 597, 9

\bibitem{Freedman}
W.L. Freedman et al., ApJ, 553, 47, 2001

\bibitem{H0lens}
L.L.R. Williams, P. Saha, AJ, 119, 439, 2000;
V.F. Cardone, S. Capozziello, V. Re, E. Piedipalumbo, A\&A, 379, 72, 2001;
V.F. Cardone, S. Capozziello, V. Re, E. Piedipalumbo, A\&A, 382, 792, 2002;
C. Tortora, E. Piedipalumbo, V.F. Cardone, MNRAS, 354, 353, 2004;
T. York, I.W.A. Browne, O. Wucknitz, J.E. Skelton, MNRAS, 357, 124, 2005

\bibitem{H0SZ}
J.P. Hughes, M. Birkinshaw, ApJ, 501, 1, 1998;
R. Saunders et al., MNRAS, 341, 937, 2003;
R.W. Schmidt, S.W. Allen, A.C. Fabian, MNRAS, 352, 1413, 2004

\bibitem{WM04}
Y. Wang, P. Mukherjee, ApJ, 606, 654, 2004

\bibitem{WT04}
Y. Wang, M. Tegmark, Phys. Rev. Lett., 92, 241302, 2004

\bibitem{HS96}
W. Hu, N. Sugiyama, ApJ, 471, 542, 1996

\bibitem{Kirk}
D. Kirkman, D. Tyler, N. Suzuki, J.M. O'Meara, D. Lubin, ApJS, 149, 1, 2003

\bibitem{Eis05}
D. Eisenstein et al., astro\,-\,ph/0501171, 2005

\bibitem{SDSSMain}
M.A. Strauss et al., AJ, 124, 1810

\bibitem{Krauss}
L. Krauss and B. Chaboyer, Science, Jan 3 2003 issue

\bibitem{Cayrel}
R. Cayrel et al., Nature, 409, 691, 2001

\bibitem{DorLil01}
M. Doran, M. Lilley, J. Schwindt, C. Wetterich, ApJ, 559, 501, 2001;

\bibitem{DorLil02}
M. Doran, M. Lilley, MNRAS, 330, 965, 2002

\bibitem{Guo}
Z.K. Guo, N. Ohta, Y.Z. Zhang, astro\,-\,ph/0505253, 2005

\bibitem{rec}
A.A. Starobinski, JETP Lett., 68, 757, 1988;
D. Huterer, M.S. Turner, Phys. Rev. D, 60, 081301, 1999;
T. Chiba, T. Nakamura, Phys. Rev. D, 62, 121301, 2000

\bibitem{exp}
P.G. Ferreira, M. Joyce, Phys. Rev. D, 58, 023503, 1998;
T. Barreiro, E.J. Copeland, N.J. Nunes, Phys. Rev. D, 61, 127301, 2000

\bibitem{RS}
C. Rubano, P. Scudellaro, Gen. Rel. Grav., 34, 307, 2001; 
M. Pavlov, C. Rubano, M. Sazhin, P. Scudellaro, ApJ, 566, 619, 2002;
C. Rubano, M. Sereno, MNRAS, 335, 30, 2002;
C. Rubano, P. Scudellaro, E. Piedipalumbo, S. Capozziello, M. Capone, Phys. Rev. D, 69, 103510, 2004;
M. Demianski, E. Piedipalumbo, C. Rubano, C. Tortora, A\&A, 431, 27, 2005

\bibitem{RP}
B. Ratra, P.J.E. Peebles, Phys. Rev. D, 37, 3406, 1988;
C. Wetterich, Nucl. Phys. B, 302, 668, 1988 

\bibitem{SUGRA}
P. Brax, J. Martin, Phys. Lett. B, 468, 40, 1999;
P. Brax, J. Martin, A. Riazuelo, Phys. Rev. D, 62, 103505, 2000;
P. Brax, J. Martin, Phys. Rev. D, 71, 063530, 2005

\end{thebibliography}
\end{document}